\documentclass[tighten,twocolumn]{aastex701}

\usepackage{graphicx}
\usepackage{ulem}
\usepackage{amsmath}
\usepackage{wrapfig}
\usepackage[thinlines]{easytable}
\usepackage{rotating}
\usepackage{tikz}
\usepackage{booktabs,multirow}

\newcommand{\msun}{{M_\odot}}

\newcommand{\nbh}{{N_{\rm{BH}}}}
\newcommand{\mbh}{{M_{\rm{BH}}}}

\graphicspath{{./}{figures/}}

\begin{document}
\shorttitle{Connecting Cores and Black Holes Across Scales}
\shortauthors{Kremer et al.}

\title{Connecting Cores and Black Hole Dynamics Across Scales: \\ From Globular Clusters to Massive Ellipticals}

\correspondingauthor{Kyle Kremer}
\author[0000-0002-4086-3180]{Kyle Kremer}
\affiliation{Department of Astronomy \& Astrophysics, University of California, San Diego; La Jolla, CA 92093, USA}
\email[show]{kykremer@ucsd.edu}

\author[0000-0002-9660-9085]{Newlin C. Weatherford}
\affiliation{The Observatories of the Carnegie Institution for Science, Pasadena, CA 91101, USA}
\email{nweatherford@carnegiescience.edu}

\author[0000-0003-3729-1684]{Philip F. Hopkins}
\affiliation{TAPIR, MC 350-17, California Institute of Technology, Pasadena, CA 91125, USA}
\email{phopkins@caltech.edu}

\author[0000-0002-1884-3992]{Nicholas Z. Rui}
\altaffiliation{NASA Hubble Fellow}
\affiliation{Department of Astrophysical Sciences, Princeton University, 4 Ivy Lane, Princeton, NJ 08544, USA}
\affiliation{Center for Interdisciplinary Exploration and Research in Astrophysics (CIERA), Northwestern University, 1800 Sherman Ave., Evanston, IL 60201, USA}
\affiliation{TAPIR, MC 350-17, California Institute of Technology, Pasadena, CA 91125, USA}
\email{nrui@princeton.edu}

\author[0000-0001-9582-881X]{Claire S. Ye}
\affil{Canadian Institute for Theoretical Astrophysics, University of Toronto, 60 St. George Street, Toronto, ON M5S 3H8, Canada}
\email{claireshiye@cita.utoronto.ca}

\begin{abstract}
The centers of massive elliptical galaxies exhibit a wide range in density profiles, from central cusps to resolved cores with order kiloparsec sizes. The cored ellipticals have been linked to the presence of supermassive black hole binaries that excavate their hosts' central stellar populations through three-body encounters. This connection between cores and black holes similarly operates in globular clusters, which also exhibit a bimodality in cored and core-collapsed architectures, respectively rich and depleted in stellar black holes. We report new estimates of the total black hole mass in 25 Galactic globular clusters based on a suite of roughly 150 Monte Carlo $N$-body simulations that fit observed surface brightness and velocity dispersion profiles. We show that both globular clusters and massive elliptical galaxies individually exhibit strong correlations between total black hole mass ($M_\bullet$) and core radius ($r_c$), and that these individual relations share a common power-law exponent to within $1\sigma$ statistical precision: $M_\bullet \sim r_c^{1.3}$. The individual relations appear to be offset, suggesting swarms of stellar black holes scour globular cluster cores more efficiently than lone supermassive black holes scour the cores of massive ellipticals. Yet the shared basis of core scouring via black hole binaries hints at a unified $M_{\bullet}-r_c$ connection across over 10 orders of magnitude in $M_\bullet$. Our findings imply core radius measurements may offer a powerful observational constraint on black hole merger rates, from kilohertz sources detectable by LIGO/Virgo/KAGRA formed in globular clusters to millihertz and nanohertz sources formed in massive elliptical galaxies.
%\vspace{1cm}
\end{abstract}

\section{Introduction}
\label{sec:intro}

The existence of nearly constant density cores appears ubiquitous across a range of astronomical systems, from dark matter halos \citep[e.g.,][]{Burkert1995} to massive elliptical galaxies \citep[e.g.,][]{Faber1997} to globular clusters \citep[e.g.,][]{Trager1995}. In all of these examples, the counterparts of cores (i.e., ``core-collapsed'' profiles or ``cusps'') are also observed, suggesting the presence of a core (or lack thereof) requires a physical explanation.

In massive elliptical galaxies, the presence of cores has been connected to supermassive black hole binaries and mergers \citep{MilosavljevicMerritt2001,Merritt2006,KormendyBender2009,Thomas2016}. As a massive black hole pair hardens via three-body interactions with stars, stars are flung from their host at high velocities, gradually depleting the central stellar population. This ``scouring'' process produces a measurable depletion in the central surface brightness, evidencing itself as a core. Previous studies have demonstrated a clear correlation between the central black hole mass and core radius, and have argued that galaxies with oversized cores are home to current (or recently merged) massive black hole binaries \citep[e.g.,][]{Begelman1980,Ebisuzaki1991,QuinlanHernquist1997,Rantala2018,Dosopoulou2021}. As these massive black hole binaries harden and ultimately merge, they are expected to be prime sources of gravitational wave emission at millihertz frequencies and below, detectable by pulsar timing arrays \citep{IPTA2022,NANOGrav2023} and the upcoming space-based interferometer LISA \citep{LISA2023}. 

\begin{figure*}
    \centering  \includegraphics[width=\linewidth]{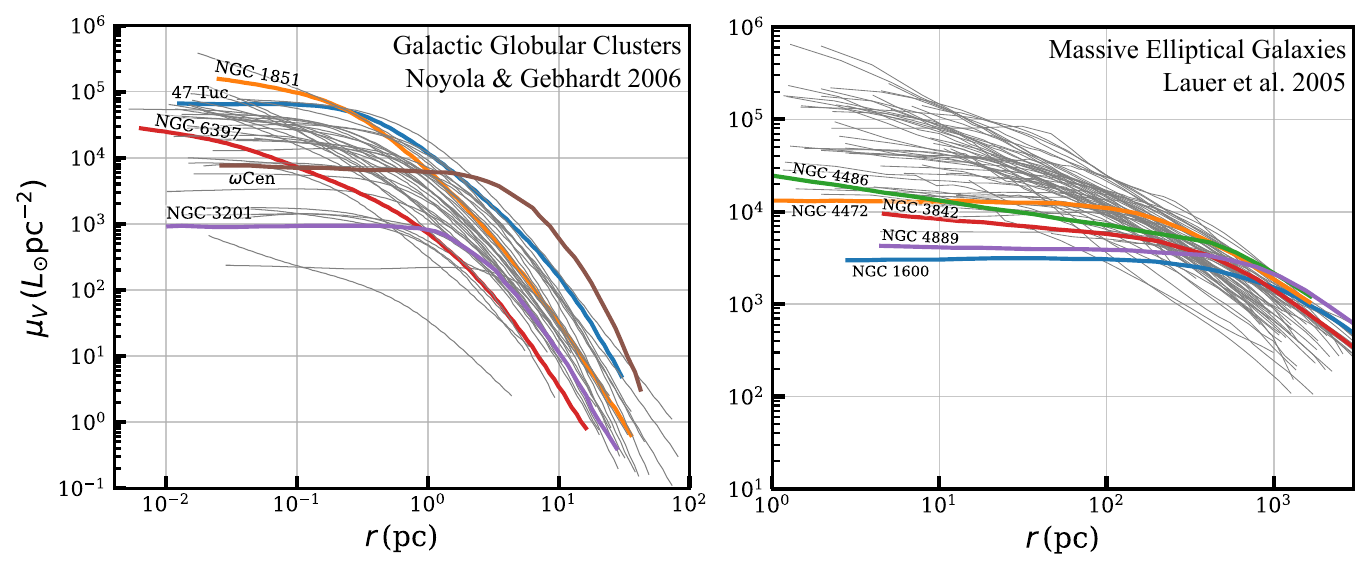}
    \caption{\textit{Left panel:} Surface brightness profiles (V-band) for Galactic globular clusters from \citet{NoyolaGebhardt2006}, illustrating the diversity of cluster core sizes. We highlight five particular clusters that are discussed in further detail in the text. \textit{Right panel:} Surface brightness profiles (also V-band) for a sample of core and coreless elliptical galaxies out to a distance of $100\,$Mpc from \citet{Lauer2005}. We highlight the brightest galaxies in the Leo Cluster (NGC 3842), Coma Cluster (NGC 4889), and Virgo Cluster (NGC 4472 as well as NGC 4486). We also highlight NGC 1600 from \citet{Thomas2016}, which has one of the faintest stellar cores of any galaxy with a black hole mass measurement.}
    \label{fig:SBPs}
\end{figure*}

Observationally, this black hole--core connection is difficult to constrain. Although central black hole masses as low as $10^5\,M_{\odot}$ can now be measured via various techniques \citep[e.g.,][]{KormendyHo2013,Peterson2014,Greene2020} the presence of an active galactic nucleus (AGN) in many of these systems generally overwhelms the inner stellar luminosity and inhibits measurement of a core, if present. In this case the black hole--core connection is generally limited to massive and quiescent galaxies with black hole masses inferred via dynamical techniques and central surface brightnesses large enough to permit a core measurement.

In the more local Universe, Galactic globular clusters present an alternate environment where the presence of cores has been well-studied. Roughly $80\%$ of Milky Way globular clusters exhibit brightness profiles with well-resolved cores of order a parsec in size \citep{Trager1995}, with the remaining 20\% exhibiting so-called core-collapsed architectures with surface brightness profiles observed to increase down to scales of $0.1\,$pc or less (see Figure~\ref{fig:SBPs}). As self-gravitating collisional systems, globular clusters feature a natural flow of energy from their dynamically ``hot'' centers to their dynamically ``cold'' outer regions, with cluster core collapse as the inescapable long-term outcome \citep{Antonov1962,LyndenBell1968,Spitzer1987}. The fact that most Galactic globular clusters are observed to be cored, suggests that some physical process acts to halt, or at least slow, the otherwise inevitable collapse.
In recent years, stellar-mass black holes have emerged as a likely explanation \citep{Merritt2004,Mackey2007,BreenHeggie2013,Kremer2020_bhburning}, motivated in part by the growing number of stellar-mass black hole binaries observed in Galactic globular clusters via both radial velocity and X-ray/radio measurements \citep{Strader2012,Chomiuk2013,MillerJones2015,Giesers2018,Giesers2019}.

\begin{figure*}
    \centering  \includegraphics[width=0.8\linewidth]{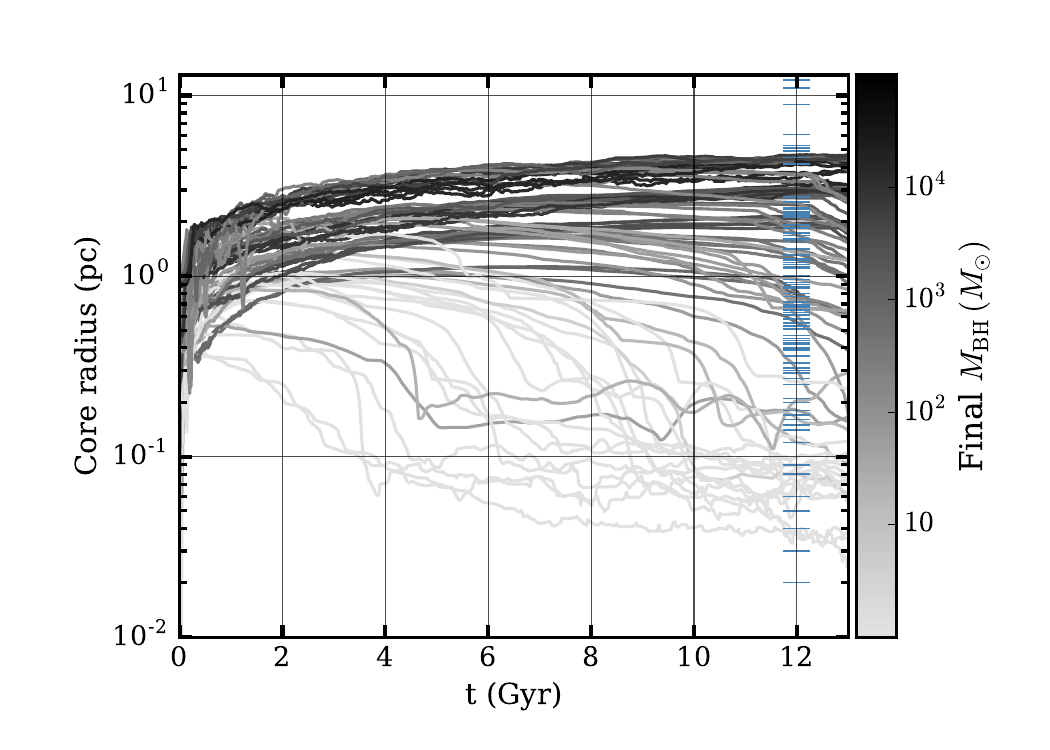}
    \caption{Time evolution of the core radius---using the density-weighted definition of \cite{CasertanoHut1985}---for all globular cluster simulations computed by \citet{Kremer2020}. Curve color denotes the total mass in black holes at the end of each simulation ($t=13\,$Gyr). As shown, clusters with larger core radius generally host more-massive black hole populations. Blue ticks on the right hand side denote present-day core radii of Milky Way globular clusters from \citet{BaumgardtHilker2018}.}
    \label{fig:CMC_rc}
\end{figure*}

Stellar-mass black hole binaries are expected to form in globular clusters via a combination of primordial pairing at the time of star formation and three-body binary encounters \citep[e.g.,][]{Kulkarni1993,Morscher2015}. Once binaries are formed, binary hardening via subsequent three-body encounters is the primary mechanism through which black holes inject energy into the stellar bulk of a globular cluster \citep{Sigurdsson1993}. The dynamical encounters that harden a binary naturally impart dynamical recoil kicks to both the interloping objects (both stars and other black holes) and the center-of-mass of the binary itself \citep{HeggieHut2003}. Black holes that are kicked out of the cluster core via this process will naturally mass-segregate back to the center via interactions with lower-mass stars in the outer regions of the host; this mass segregation further injects energy into the stellar bulk. The cumulative effect of many repeated iterations of this process---hardening, recoil, and mass segregation---is energy flow from the cluster's center (where black holes preferentially reside) to the surrounding regions. As long as a sufficiently massive black hole population remains present, a roughly constant density core is sustained \citep{Mackey2007}. Only after the black hole population has been fully depleted via dynamical ejections will the cluster undergo core collapse \citep{kremer2019initial}. This basic process has now been demonstrated in a number of ways including analytic calculations \citep{BreenHeggie2013,AntoniniGieles2020}, direct $N$-body simulations \citep{wang2016dragon,Banerjee2017}, and Monte Carlo cluster simulations \citep{Morscher2015,Kremer2018_ngc3201,Askar2018,Weatherford2020,Rodriguez2023}.

There are obvious differences between the massive elliptical regime, where a single massive black hole binary fixed at the center directly ejects many stars with extreme velocities, and the globular cluster regime, where a spatially extended population of stellar black holes gradually transfer energy to stars via many separate weaker encounters. But in both cases binary hardening via three-body encounters is the fundamental phenomenon by which black holes drive expansion of their host's core, motivating a possible connection across these regimes despite their huge range in physical scale. This Letter broadly explores this black hole--core connection. Using a large suite of globular cluster $N$-body models computed using the code \texttt{CMC}, we demonstrate that the dynamical influence of stellar-mass black holes naturally accounts for the observed range in core sizes of Galactic globular clusters (Section~\ref{sec:CMC}). In Section~\ref{sec:BHcore}, we link globular clusters to massive elliptical galaxies, demonstrating the dynamical role of black holes across stellar populations spanning nearly eleven orders of magnitude in black hole mass. In Section~\ref{sec:discussion}, we discuss our results and implications. We discuss a few future tests of this relation and describe how this black hole--core connection can be leveraged to constrain black hole mergers across gravitational wave frequencies from nanohertz to kilohertz.

\begin{figure*}
    \centering  \includegraphics[width=0.7\linewidth]{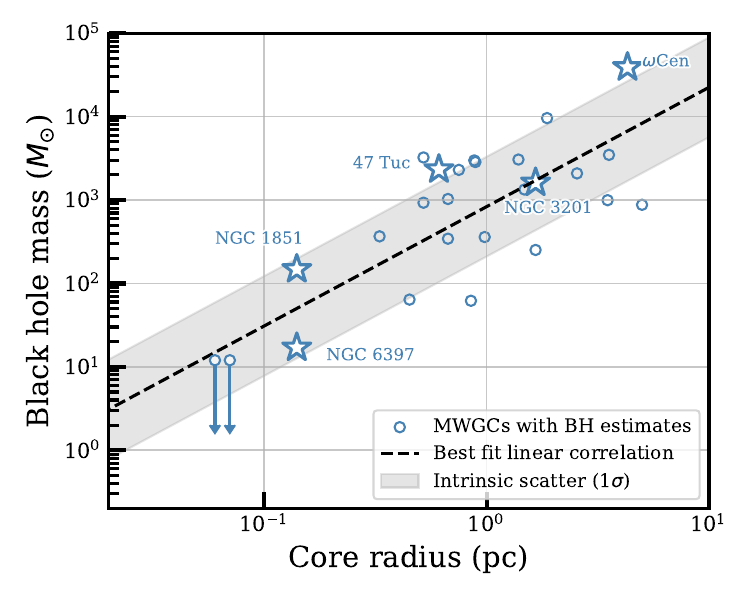}
    \caption{Total mass of the stellar black hole populations predicted for various Milky Way globular clusters versus core radius. Blue scatter points show the black hole predictions using the methods of this paper (see Table~\ref{table:nbh_mbh}), with stars highlighting a few specific clusters discussed further in the text. For the core-collapsed clusters NGC~6293 and NGC~6681, we plot the mass of the final black hole ejected from the best-fit simulations ($12\,M_{\odot}$), with the arrows indicating these clusters may host zero black holes at present. All core radius measurements are taken from \citet{BaumgardtHilker2018}. The dashed black curve shows our best-fit linear correlation $\log_{10} (M_{\bullet}/M_{\odot}) = (2.92 \pm 0.12) + (1.43 \pm 0.23 )\log_{10}(r_c/\rm{pc})$.}
    \label{fig:Mbh_rc_GC}
\end{figure*}

\section{Globular Cluster Simulations}
\label{sec:CMC}

To study the role of black hole binaries in globular clusters, we use a suite of cluster simulations computed using the \texttt{Cluster Monte Carlo} code \texttt{CMC}, a fully-parallelized H\'{e}non-type Monte Carlo code for modeling the evolution of globular clusters. \texttt{CMC} incorporates relevant dynamical processes such as two-body relaxation and resonant encounters, as well as stellar and binary star evolution using the population synthesis code \texttt{COSMIC} \citep{Breivik2020}. Additionally, \texttt{CMC} output can be converted into various observational quantities, such as surface brightness profiles, velocity-dispersion profiles, and color-magnitude diagrams, using the \texttt{cmctoolkit} package \citep{Rui2021}, enabling robust comparisons with realistic globular clusters. This has enabled detailed study of a range of compact object sources including radio pulsars \citep{Ye2019}, X-ray binaries \citep{Kremer2018a}, white dwarfs \citep{Kremer2021_wd}, and binary black hole mergers \citep{Rodriguez2016a}. For a detailed review of the methods of \texttt{CMC}, see \citet{Rodriguez2022}. Here we use the simulations published as part of the \texttt{CMC Cluster Catalog} \citep{Kremer2020}. In aggregate, this suite of 148 independent models is designed as a proxy for the current Galactic globular cluster sample, effectively spanning the observed ranges in cluster mass, core and half-light radii, metallicity, and Galactocentric position. Hereafter, when reporting core radius from \texttt{CMC} simulations, we use the density-weighted definition of \cite{CasertanoHut1985}.

\begin{figure*}
    \centering  \includegraphics[width=0.7\linewidth]{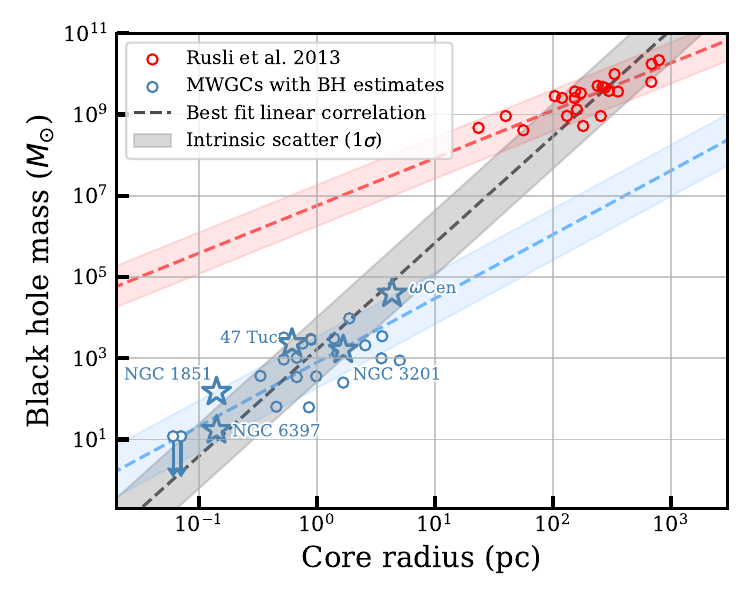}
    \caption{Black hole mass versus core radius across stellar populations. In red we show massive elliptical galaxies from \citet{Rusli2013} with resolved cores and dynamical black hole mass measurements. The red curve shows the best-fit linear correlation to these data from \citet{Thomas2016}. In blue we show predicted masses of stellar-mass black hole populations for a number of Galactic globular clusters, with the best-fit linear correlation from Figure~\ref{fig:Mbh_rc_GC} shown again here in blue. The slopes of these two populations are consistent to within $1\sigma$, despite spanning orders of magnitude across both axes. The dashed black curve shows our best-fit global correlation for globular clusters and massive ellipticals combined.}
    \label{fig:Mbh_rc}
\end{figure*}

\subsection{How Black Holes Shape Globular Cluster Cores}

In Figure~\ref{fig:CMC_rc}, we show the cluster core radius evolving with time for all simulations in the \texttt{CMC Cluster Catalog} that (1) have metallicity $Z\leq 0.1\,Z_{\odot}$ representative of old and metal-poor Galactic globular clusters and (2) survive for at least $13\,$Gyr (76 models total). The color of each curve indicates the total mass of the black hole population at the end of the simulation ($t=13\,$Gyr), as shown in the color bar. The blue ticks on the right-hand side mark the observed core radii \citep[from][]{BaumgardtHilker2018} for all Galactic globular clusters with masses of at least $5\times10^4\,M_{\odot}$. 

At early times ($t<100\,$Myr), core expansion is dominated by mass loss of massive stars \citep[e.g.,][]{PortegiesZwart2010}. At late times highlighted in this figure, the evolution of the core is dominated by the presence (or lack thereof) of black holes in the cluster center \citep{Mackey2007,BreenHeggie2013,Morscher2015,Kremer2018_ngc3201,kremer2019initial}. As shown, cluster simulations that have lost nearly all of their black holes undergo core collapse by $10\,$Gyr ($r_c \lesssim0.1$pc), while simulations with total black hole mass ${\gtrsim}10^3\,M_{\odot}$ still contain large cores of order a parsec or larger. Cumulatively, this effect reproduces well the complete range in core radii observed for Galactic clusters. 

\subsection{Estimating Black Hole Populations in Globular Clusters}

These models can be also used to predict the total number and mass of the black hole populations in specific Galactic globular clusters at present.
In \citet{Rui2021}, we implemented a $\chi^2$ statistic based upon matches to observed surface brightness and velocity dispersion profiles to measure goodness-of-fit of a given model cluster to a given observed cluster. We
showed that a number of Galactic clusters are well-matched by
at least one \texttt{CMC Cluster Catalog} model, enabling us to predict the black hole populations in these particular systems. In addition to 21 clusters with good model fits in the \texttt{CMC Cluster Catalog}, we also include here black hole predictions from four additional clusters with good-fitting \texttt{CMC} models published separately: for NGC~3201, we use the models of \citet{kremer2019initial}, for 47 Tucanae, we use the models of \citet{Ye2022_47tuc}, and for the core-collapsed clusters NGC~1851 and NGC~6397, we use the models of \citet{Ye2024_massgap} and \citet{Kremer2021_wd}, respectively. In addition to matching observed surface brightness and velocity dispersion profiles, these various best-fit models have been shown to match well the various compact object sources in relevant clusters, including millisecond pulsars, cataclysmic variables, and low-mass X-ray binaries.

We list the predicted total number and mass of black hole populations for the 25 Galactic globular clusters with good model fits in Table~\ref{table:nbh_mbh} of the Appendix. We also show the our computed surface brightness and velocity dispersion profiles for all best-fit models compared to the observed profiles in Figures~\ref{fig:sbp_vdp_set_1}--\ref{fig:sbp_vdp_set_4}, 

A number of previous numerical studies have estimated the black hole populations in a number of Galactic globular clusters using a variety of methods \citep{Mackey2007,HeggieGiersz2014,Kremer2018_ngc3201,kremer2019initial,Askar2018,Weatherford2018,Weatherford2020,Vitral2022,Dickson2024,DellaCroce2024}. Although the exact number of black holes predicted varies from study to study, all studies predict a similar correlation between the size of the black hole population and core radius.

\section{The Black Hole--Core Connection Across Scales}
\label{sec:BHcore}

\subsection{Globular Clusters}

Figure~\ref{fig:Mbh_rc_GC} shows in blue the total mass of the stellar-mass black hole populations ($M_\bullet$) predicted for the Galactic globular clusters in Table~\ref{table:nbh_mbh} versus observed core radius. We highlight as blue stars a handful of notable clusters: NGC~3201 (predicted to host $M_{\bullet} \approx 1615\,M_{\odot}$) contains \textit{three} observed black hole candidates, identified via the radial velocity variation of their luminous binary companions \citep{Giesers2018}. This cluster is now well-established to host a robust black hole population and has been studied at length in a number of prior studies \citep[e.g.,][]{Kremer2018_ngc3201,Giesers2019}. 47~Tucanae (predicted to host $M_{\bullet} \approx 2330\,M_{\odot}$) is one of the most massive and dense Galactic globular clusters and has long been associated with a central intermediate-mass black hole and/or a population of stellar-mass black holes \citep{Kiziltan2017,Weatherford2018,Weatherford2020, Abbate2018,HenaultBrunet2020,Ye2022_47tuc,Paduano2024}. 47~Tucanae also hosts the only ultracompact black hole--white dwarf binary candidate in the Milky Way \citep{MillerJones2015,Bahramian2017}. NGC~6397 (predicted to host $M_{\bullet} \approx 17\,M_{\odot}$) is the closest (to Earth) core-collapsed globular cluster in the Milky Way. Recent studies \citep{Rui2021RNAAS,Kremer2021_wd,Vitral2022} have demonstrated this cluster likely contains a large central population of massive white dwarfs and neutron stars. NGC~1851 (predicted to host $M_{\bullet} \approx 148\,M_{\odot}$) is a relatively massive core-collapsed cluster, also expected to host a large central white dwarf and neutron star population \citep{Ye2024_massgap}. Recent radio observations identified a mass-gap black hole candidate that may be evidence for ongoing binary neutron star mergers in this cluster \citep{Barr2024}. 
Finally, $\omega$~Centauri ($\omega$~Cen) is the most massive globular cluster in the Galaxy, long thought to be the core of a stripped dwarf galaxy accreted by the Milky Way during a previous merger. \citet{Haberle2024} recently inferred the presence of a central intermediate-mass black hole in $\omega$~Cen of mass ${\sim}10^4\,M_{\odot}$, with a firm lower limit of $8200\,M_{\odot}$, based on proper motion measurements (with Hubble Space Telescope) of seven fast-moving stars near the cluster's center. Recent \texttt{CMC} simulations of $\omega$~Cen predict the growth of this massive black hole is primarily driven by mergers with stellar-mass black holes over the cluster lifetime \citep{GonzalezPrieto2025}. In Figure~\ref{fig:Mbh_rc_GC}, we adopt $M_{\bullet} = 3\times10^4\,M_{\odot}$ for $\omega$~Cen, as predicted by \cite{Haberle2024}.

We fit the data in Figure~\ref{fig:Mbh_rc_GC} using a least-squares regression. Our best-fit linear correlation is $\log_{10} (M_{\bullet}/M_{\odot}) = (2.92 \pm 0.12) + (1.43 \pm 0.23 )\log_{10}(r_c/\rm{pc})$. We find an intrinsic scatter (1$\sigma$) of $0.60\,$dex for this relation. The black dashed curve in Figure~\ref{fig:Mbh_rc_GC} shows the best-fit correlation. The exponent of this power law is consistent with the correlation between $M_\bullet/M_{\rm cl}$ and $r_c/r_{\rm h}$ derived from mass segregation arguments in \citet{Weatherford2020}---their Figure~7---which has a power law exponent roughly $1.5$, where $M_{\rm cl}$ and $r_{\rm h}$ are the cluster's mass and half-light radius, respectively. For completeness, we also include in the Appendix Figure~\ref{fig:GC_appendix}, which shows (the weaker) correlations of our predicted black hole masses with other cluster parameters. Figure~\ref{fig:GC_appendix} also demonstrates that our 25 clusters with predicted black hole populations are collectively representative of the full Milky Way globular cluster parameter space.

\begin{figure*}  
\centering  \includegraphics[width=\linewidth]{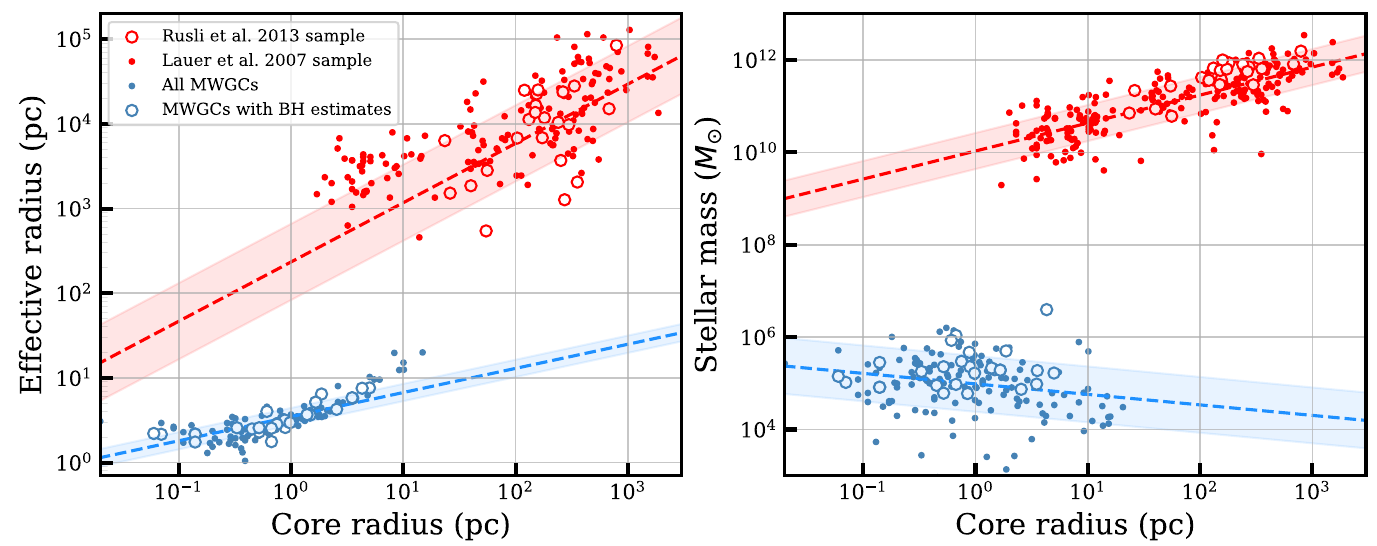}
    \caption{Effective radius versus core radius (\textit{left panel}) and total stellar mass versus core radius (\textit{right panel}) for Galactic globular clusters (blue) and massive elliptical galaxies (red). Filled blue circles show the full population of Galactic globular clusters; open blue circles show those in Figure~\ref{fig:Mbh_rc} with black hole mass estimates. Filled red circles show the massive early-type galaxies cataloged by \citet{Lauer2007}; open circles show those with black hole mass and core measurements by \cite{Rusli2013}---also shown in Figure~\ref{fig:Mbh_rc}. Blue and red bands show the best-fit local linear correlations for globular clusters and massive ellipticals, respectively.}
    \label{fig:Reff_rc}
\end{figure*}

\subsection{Connecting to Massive Elliptical Galaxies}

We demonstrate the black hole--core radius correlation across scales in Figure~\ref{fig:Mbh_rc}. Blue points again show our predictions for Galactic globular clusters, with the blue curve showing the best-fit linear correlation from Figure~\ref{fig:Mbh_rc_GC}. In red, we plot values for massive elliptical galaxies published in \citet{Rusli2013}, with the red curve showing the best-fit for these systems alone: $\log_{10} (M_{\bullet}/M_{\odot}) = (10.27\pm0.51) + (1.17\pm0.14)\log_{10}(r_c/\rm{kpc})$ \citep[see][]{Thomas2016}. \textit{We note the striking agreement in the slopes of these two best-fit curves (consistent to within $1\sigma$), despite the very different scales of the two populations.}

It is expected that the local relations for globular clusters and massive elliptical galaxies are offset vertically, as these are clearly distinct populations from a stellar dynamics perspective. Globular clusters have much shorter relaxation times and, although three-body encounters are the underlying driving mechanism by which the black holes influence the stellar bulk in both populations, the details are quite different. In the supermassive black hole regime, the black hole binary remains effectively stationary at the host galaxy's center, and objects interacting with the binary are ejected from the galaxy entirely \citep[e.g.,][]{MilosavljevicMerritt2001}. In this regime, the massive black hole binary scours the core by depleting the luminous stellar population, a far more extreme version of the more gradual heating from a population of stellar-mass black holes. Globular clusters do eventually eject their stellar-mass black holes entirely from the cluster, but typically only after many partial ejections from the cluster's core to its halo. Thus mass segregation plays an important role in the energy exchange as the black holes act in opposition to core collapse \citep{Merritt2004}. The vertical offset of the two curves could be interpreted to arise from the relative efficiency of core-scouring of the two populations. For example core radii of roughly $10\,$pc are supported by $M_{\bullet} \approx 10^4\,M_{\odot}$ for globular clusters, while the same core radius requires $M_{\bullet} \approx10^8\,M_{\odot}$ for massive ellipticals. In other words, although the underlying physical process is similar, the dense black hole subsystems of globular clusters scour a similar volume with substantially smaller mass relative to supermassive black hole binaries in elliptical galaxies.

For completeness, we also fit the combined cluster+elliptical galaxy dataset and identify a global correlation $\log_{10} (M_{\bullet}/M_{\odot}) = (3.22 \pm 0.14) + (2.62 \pm 0.09 )\log_{10}(r_c/\rm{pc})$, with intrinsic scatter of $0.80\,$dex.\footnote{Note that the globular cluster and massive elliptical samples of Figure~\ref{fig:Mbh_rc} contain 25 and 21 data points, respectively. Thus the best-fit global relation features roughly equal weights from each of two these populations.} This is shown as the black curve in Figure~\ref{fig:Mbh_rc}. As shown, the global fit is slightly steeper than the local fits for globular clusters and massive ellipticals alone. In this sense, the global fit is not intended to argue that massive ellipticals are merely ``scaled up'' globular clusters. Rather, this global trend could be interpreted as the transition in scouring efficiency from diffuse black hole subsystems to single massive black hole binaries. Future observations (Section~\ref{sec:discussion}) may test whether this global fit holds physical significance.

\subsection{Are Systems with Larger Cores Simply Larger?}
\label{sec:larger_systems_larger}

There are also well-known empirical relations between black hole mass and other stellar properties---for instance, more massive populations host more massive black holes \citep[e.g.,][]{KormendyRichstone1995,Magorrian1998}. So at some level, the overall trends between $M_{\bullet}$ and $r_c$ in Figure~\ref{fig:Mbh_rc} stem from more-massive populations simply having larger core radii. We explore this point in Figure~\ref{fig:Reff_rc}, where we plot effective radius $R_{\rm eff}$ (left panel) and total stellar mass (right panel) versus core radius for all Galactic globular clusters (blue) and massive ellipticals (red) cataloged by \citet{Lauer2007}. Open blue circles denote the globular clusters in Figure~\ref{fig:Mbh_rc} for which we estimate black hole population masses. Open red circles denote the galaxies shown in Figure~\ref{fig:Mbh_rc} with both black hole and core radius measurements \citep{Rusli2013}.

We again show the best-fit local linear correlations for each of these populations (for consistency with Figure~\ref{fig:Mbh_rc}, fitting to only the systems with black hole mass measurements; open circles). We can again compare the local slopes for globular clusters ($\log_{10} [R_{\rm eff}/\rm{pc}] \propto [0.29 \pm 0.05]\log_{10}[\it{r_c}/\rm{pc}]$ and $\log_{10} [M_{\star}/M_{\odot}] \propto [-0.23 \pm 0.08 ]\log_{10}[\it{r_c}/\rm{pc}]$) to the local slopes for massive ellipticals ($\log_{10} [R_{\rm eff}/\rm{pc}] \propto [0.70 \pm 0.25]\log_{10}[\it{r_c}/\rm{pc}]$ and $\log_{10} [M_{\star}/M_{\odot}] \propto [0.61 \pm 0.03 ]\log_{10}[\it{r_c}/\rm{pc}]$) in each of these panels. Unlike in Figure~\ref{fig:Mbh_rc}, the respective local slopes differ substantially (by at least roughly $2\sigma$), reiterating that the $M_{\bullet}$--$r_c$ correlation holds unique physical significance across these populations and scales.

In a global sense, the relatively massive and extended elliptical galaxies do of course have intrinsically larger core radii compared to globular clusters. However, the global $M_{\bullet}$--$r_c$ correlation (black curve in Figure~\ref{fig:Mbh_rc}) is stronger than the global correlations for $R_{\rm eff}$--$r_c$ and $M_{\star}-r_c$ of Figure~\ref{fig:Reff_rc}. Again combining the globular cluster and elliptical datasets, we can evaluate the strength of each global correlation via Pearson correlation coefficients $r$, finding $r=0.97$ for $M_{\bullet}$--$r_c$, $r=0.89$ for $R_{\rm eff}$--$r_c$, and $r=0.81$ for $M_{\star}$--$r_c$. This further suggests that dynamical coupling between black holes and cores is physical meaningful relative to other quantities.

Comparison of Figures~\ref{fig:Mbh_rc} and \ref{fig:Reff_rc} raises several other interesting questions. For example, although $M_{\bullet}$ is strongly correlated with $r_c$ for both globular clusters and massive ellipticals, the right panel of Figure~\ref{fig:Reff_rc} shows that stellar mass is correlated with $r_c$ for ellipticals, but not for globular clusters. This is consistent with well-known relations between black hole mass and stellar mass for massive ellipticals \citep[e.g.,][]{KormendyHo2013,McConnellMa2013}, but it also suggests that the black hole mass fraction is highly variable across globular clusters. Indeed, a similar point is also illustrated in the top panels of Figure~\ref{fig:GC_appendix}, where significant scatter is evident in relations showing $M_{\bullet}$ versus $r_h$, $M_{\star}$, and $\sigma$, compared to the relatively tight $M_{\bullet}-r_c$ relation shown in Figure~\ref{fig:Mbh_rc_GC}. In globular clusters, structural properties are significantly influenced by the systems' tendency to dynamically relax. Observed clusters exhibit a broad range of relaxation times \citep[e.g.,][]{BaumgardtHilker2018}, in turn leading to significant scatter. This is not as relevant for massive ellipticals, which have significantly longer relaxation times. Thus, in the elliptical regime, these properties connect more directly to the growth history of the systems, which is well understood through many previous studies to result in strong correlation between quantities such as $M_{\bullet}$, $M_{\star}$, $R_{\rm eff}$, etc. The result that there \textit{is} a clear correlation between $M_{\bullet}$ and $r_c$ in globular clusters demonstrates the dominant role of black holes in shaping the central dynamics of the system. It is expected that the black holes would exhibit the largest influence on the core, simply because this is where black holes in general spatially reside \citep[e.g.,][]{Morscher2015,wang2016dragon,kremer2020modeling}.

\section{Summary \& Discussion}
\label{sec:discussion}

To summarize, we have reached the following key conclusions:

\begin{itemize}
    \item By matching globular cluster simulations to observations, we have estimated the number and total mass of stellar-mass black hole populations in 25 different Milky Way globular clusters. Predicted black hole populations range in number from zero to several hundred. 
    \item Prior studies have demonstrated the dynamics of black holes in the core of a globular cluster play a key role in the evolution of the cluster's structural profile. We leverage our predicted black hole population to show that Milky Way globular clusters exhibit a clear correlation between the total mass of the present-day black hole population and the cluster core radius.
    \item We explore a connection to prior studies that have demonstrated a similar link between black hole dynamics and cores in massive elliptical galaxies. We demonstrate a clear empirical correlation between the total mass of black hole populations and core radius across a wide range in stellar populations.
\end{itemize}

\subsection{Future Tests}
\label{sec:tests}

Outside of the Milky Way, globular cluster cores are also directly measured for a handful of systems in the Magellanic clouds \citep[e.g.,][]{Elson1989} and for a subset of clusters in M31 \citep[e.g.,][]{Barmby2002}. At distances of a megaparsec or more, resolution of parsec-sized globular cluster cores becomes difficult, although indirect measurements can be obtained via King model fitting \citep[e.g.,][]{King1962,Larsen1999}. Thirty-meter class telescopes equipped with adaptive optics should enable direct core measurements for a much larger sample of extragalactic globular clusters, potentially enabling future extension of the black hole--core relation to a large population of clusters.

Many galaxies in the local universe have black hole masses in the range $10^5$--$10^7\,M_\odot$, but measuring core radii for these systems is a challenge due to their faint surface brightness (see Section~\ref{sec:intro}). This contributes to a notable absence in Figure~\ref{fig:Mbh_rc} of observed systems with core radii of roughly $10$--$50\,$pc. Measurement of cores for elliptical systems in this regime may be an ideal science task for JWST, as the contribution of AGN emission in these systems is expected to be less prominent in the infrared. Future measurements of the cores for such galaxies would test the empirical trend shown in Figure~\ref{fig:Mbh_rc}. Note, however, that this trend is likely to be strongest for old and gas-poor galaxies with massive black hole binaries assembled in gas-poor galaxy mergers. The stellar population formed in a compact central starburst during a gas-rich merger could likely allow a cusp and black hole binary to be present simultaneously \citep[e.g.,][]{Hopkins2008,Hopkins2009}.

\subsection{Measured Cores as Constraints on Black Hole Merger Rates}

This dynamical connection between black hole binaries and cores suggests the possibility that core measurements may constrain black hole merger rates and associated gravitational wave sources. Indeed, previous studies have attempted to leverage this connection in the contexts of both globular clusters and galaxies. Cluster simulations hosting stellar-mass black hole populations consistent with observed core radii of the Galactic globular clusters (Figure~\ref{fig:CMC_rc}) predict binary black hole merger rates of order $10\,\rm{Gpc}^{-3}\rm{yr}^{-1}$ \citep{RodriguezLoeb2018,Kremer2020}, comparable to the rate inferred from the latest LIGO/Virgo/KAGRA (LVK) data \citep{LIGO2023}. \citet{FishbachFragione2023} demonstrated this black hole--core connection can be leveraged in the opposite direction, showing that the LVK data can place constraints on the highly uncertain birth sizes of globular clusters. Similarly, the presence of cores in massive elliptical galaxies can constrain the supermassive black hole merger rates in these systems \citep[e.g.,][]{Begelman1980,Faber1997, KormendyBender2009}.

Assuming merger occurs near the radius $R_s \sim GM_{\bullet}/c^2$, the gravitational wave frequency at merger scales as $f_{\rm GW} \sim 10^{-2}\,(M_{\bullet}/10^6\,M_{\odot})^{-1}\,$Hz. Thus black holes with masses of roughly $10^6\,M_{\odot}$ are ideal targets for millihertz detectors like LISA. In this case, future detections of elliptical galaxies with cores of order $10\,$pc (Section~\ref{sec:tests}) could provide powerful insight into the black hole merger rate in this mass regime and a unique test of the LISA merger rate.

\begin{acknowledgments}

We thank the anonymous referee for constructive comments on the interpretation of our results, which significantly improved the manuscript. N.Z.R. acknowledges support from the National Science Foundation Graduate Research Fellowship under Grant No. DGE‐1745301, from the United States--Israel Binational Science Foundation through grant BSF-2022175, and through the NASA Hubble Fellowship grant
HST-HF2-51589.001-A awarded by the Space Telescope Science Institute, which is operated by the
Association of Universities for Research in Astronomy, Inc., for NASA, under contract
NAS5-26555. C.S.Y. acknowledges support from the Natural Sciences and Engineering Research Council of Canada (NSERC) DIS-2022-568580.
\end{acknowledgments}

\begin{contribution}
This work was conceived by K.K. and fleshed out in discussions with N.C.W. and P.F.H. K.K. led the write-up with key contributions from N.C.W. and feedback from all other authors. Data analysis and presentation was split evenly between K.K. and N.C.W., and was based heavily on data from N.Z.R. \citep{Rui2021}, some previously unpublished. C.S.Y. also contributed the data on BHs in her models for NGC~104 and NGC~1851.
\end{contribution}

\bibliographystyle{aasjournal}
\bibliography{mybib}

\appendix

\setcounter{figure}{0}
\renewcommand{\thefigure}{A\arabic{figure}}

\begin{figure*}[b!]  
\centering  \includegraphics[width=\linewidth]{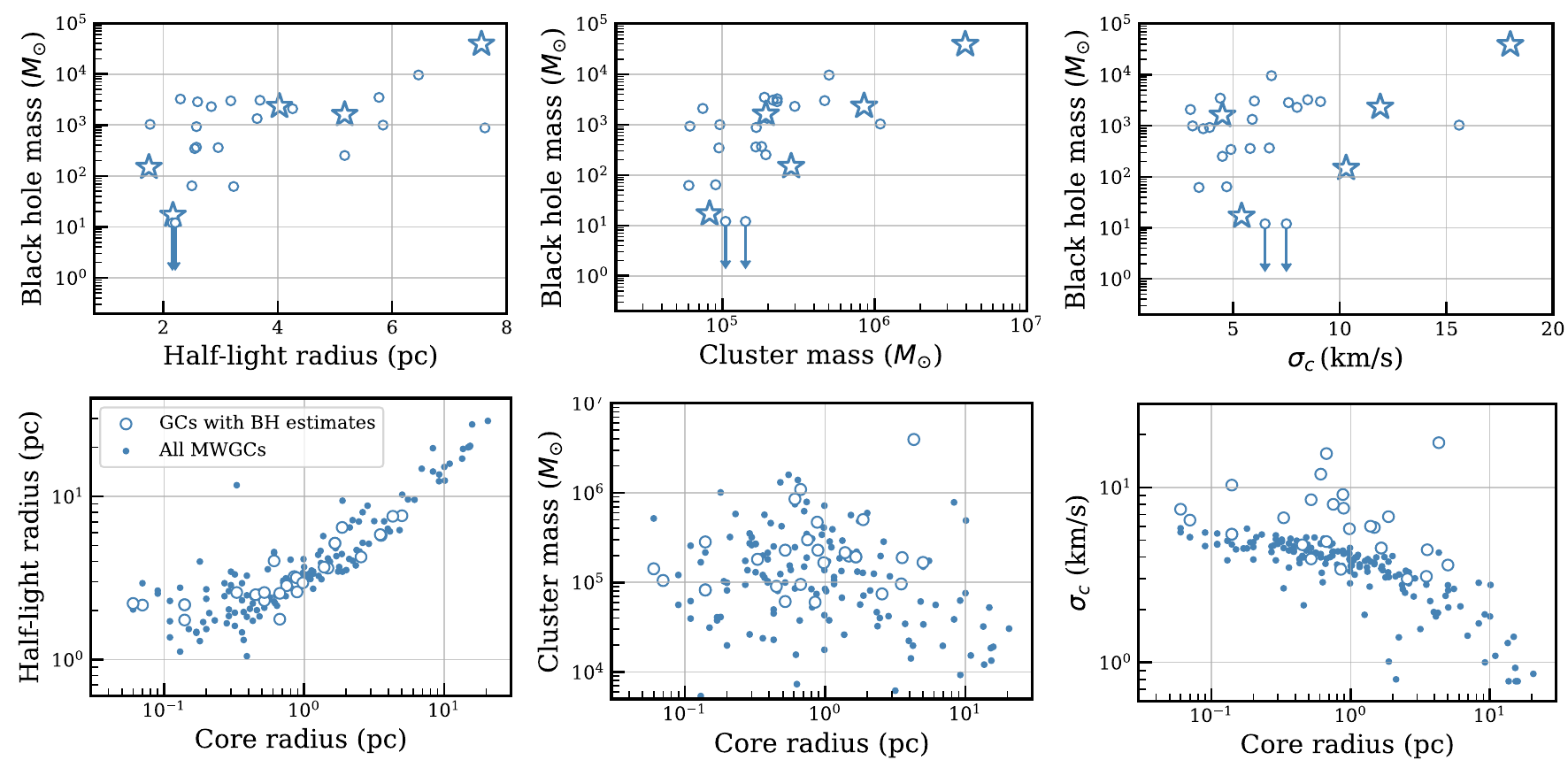}
    \caption{\textit{Top panels:} Total mass of black hole populations in our Milky Way globular clusters of interest versus half-light radius, cluster mass, and central velocity dispersion, $\sigma_c$. As in Figure \ref{fig:Mbh_rc}, open circles denote clusters from the \texttt{CMC Cluster Catalog} using the methods of \citet{Rui2021}, while stars show clusters with previously published \texttt{CMC} fits. These three datasets feature Pearson coefficients of $0.58$, $0.61$, and $0.26$, respectively (compared to $r=0.80$ for the $M_{\bullet}$--$r_c$ correlation in Figure~\ref{fig:Mbh_rc}). \textit{Bottom panels:} Half-light radius, total cluster mass, and $\sigma_c$ versus core radius for all Milky Way globular clusters \citep[data from][]{BaumgardtHilker2018}. Open circles again denote the specific clusters with black hole population estimates. Note the clusters with black hole estimates effectively span the full parameter space of the full Milky Way globular cluster population.}
    \label{fig:GC_appendix}
\end{figure*}

\setcounter{table}{0}
\renewcommand{\thetable}{A\arabic{table}}

\floattable
\begin{deluxetable*}{lc|ccccc|ccccc}
\tabletypesize{\footnotesize}
\tablewidth{0pt}
\tablecaption{Number and total mass of stellar-mass black holes in Milky Way globular clusters, as inferred from \texttt{CMC} model snapshots that fit well the cluster's surface brightness and velocity dispersion profiles \citep{Rui2021}. $N_{\rm snaps}$ denotes the total number of ``good-fit'' \texttt{CMC} models for each cluster. The quoted $N_{\rm BH}$ and $M_{\rm BH}$ percentile values are derived from these specific snapshots. The bottom four rows show black hole estimates for clusters from previously published \texttt{CMC} models (see references included). \label{table:nbh_mbh}}
\tablehead{
    \colhead{Cluster} &
    \colhead{$N_{\rm snaps}$} &
    \multicolumn{5}{c}{Percentiles in $\nbh$} &
    \multicolumn{5}{c}{Percentiles in $\mbh\ [\msun]$} \\%[-0.1cm]
    \cmidrule(lr){3-7} \cmidrule(lr){8-12}%\cline{2-6} \cline{7-11}
	\colhead{} &
    \colhead{} &
	\colhead{0\%} &
	\colhead{16\%} &
	\colhead{50\%} &
	\colhead{84\%} &
	\colhead{100\%} &
	\colhead{0\%} &
	\colhead{16\%} &
	\colhead{50\%} &
	\colhead{84\%} &
	\colhead{100\%}\vspace{-0.07cm}
}
\startdata
NGC 0288 &  16    &     48 &  54 &  70 &  79 &  88    &     679 &   775 &   996 &  1135 &  1266 \\
%NGC 1851 &   8    &     63 &  63 &  64 &  71 & 115    &     546 &   546 &   557 &   621 &  1091 \\
NGC 1904 &   7    &     24 &  25 &  30 &  35 &  35    &     293 &   306 &   367 &   434 &   434 \\
%NGC 3201 &  15    &      9 &  13 &  19 &  97 & 101    &     104 &   162 &   252 &  1443 &  1512 \\
NGC 4372 &  60    &     93 & 106 & 217 & 348 & 375    &    1378 &  1601 &  3480 &  5822 &  6334 \\
NGC 5024 &  78    &    228 & 533 & 604 & 659 & 695    &    3080 &  8360 &  9604 & 10575 & 11208 \\
NGC 5897 &  14    &     44 &  44 &  56 &  65 &  67    &     681 &   685 &   877 &  1016 &  1046 \\
NGC 5986 &  20    &    147 & 155 & 170 & 190 & 206    &    1945 &  2072 &  2299 &  2597 &  2862 \\
NGC 6121 &  16    &      1 &   1 &   5 &   6 &   6    &      13 &    13 &    64 &    75 &    75 \\
NGC 6171 &   1    &     72 &  72 &  72 &  72 &  72    &     929 &   929 &   929 &   929 &   929 \\
NGC 6293 &  59    &      0 &   0 &   0 &   0 &  16    &       0 &     0 &     0 &     0 &   192 \\
NGC 6352 &  11    &      1 &   1 &   4 &   7 &   7    &      12 &    12 &    62 &    85 &    85 \\
%NGC 6397 &  11    &      1 &   1 &   1 &   1 &   1    &      13 &    16 &    17 &    18 &    18 \\
NGC 6496 &  12    &    254 & 257 & 271 & 286 & 290    &    1955 &  1974 &  2087 &  2204 &  2233 \\
NGC 6539 &   6    &    245 & 245 & 249 & 252 & 254    &    3008 &  3008 &  3058 &  3092 &  3131 \\
NGC 6553 &  76    &    143 & 378 & 431 & 496 & 519    &    1088 &  2833 &  3240 &  3741 &  3922 \\
NGC 6569 &  25    &    198 & 219 & 233 & 249 & 254    &    2362 &  2656 &  2853 &  3064 &  3131 \\
NGC 6624 &  13    &      2 &   2 &   6 &   6 &   9    &      23 &    23 &    57 &    58 &    90 \\
NGC 6656 &  70    &    147 & 175 & 214 & 253 & 266    &    1945 &  2377 &  2985 &  3585 &  3781 \\
NGC 6681 &  49    &      0 &   0 &   0 &   0 &   0    &       0 &     0 &     0 &     0 &     0 \\
NGC 6712 &  47    &     20 &  24 &  29 & 117 & 136    &     223 &   276 &   344 &  1580 &  1851 \\
NGC 6723 &  29    &     72 &  74 & 100 & 122 & 136    &     929 &   965 &  1337 &  1642 &  1851 \\
NGC 6779 &  37    &      9 &  19 &  30 & 131 & 139    &     104 &   248 &   360 &  2005 &  2125 \\
Terzan 5 &  44    &     67 &  85 & 144 & 181 & 202    &     483 &   606 &  1029 &  1296 &  1446 \\
\hline
\hline
NGC 0104 \citep{Ye2022_47tuc} & 18 & 162 & 168 & 185 & 199 & 206 & 2061 & 2116 & 2346 & 2546 & 2650\\
NGC 1851 \citep{Ye2024_massgap} & 21 & 19 & 20 & 28 & 35 & 45 & 74 & 80 & 138 & 212 & 284\\
NGC 3201 \citep{kremer2019initial} & 4 & 109 & 111 & 119 & 131 & 137 & 1460 & 1486 & 1615 & 1814 & 1908 \\
NGC 6397 \citep{Kremer2021_wd} & 11 &      1 &   1 &   1 &   1 &   1    &      13 &    16 &    17 &    18 &    18 \\
\enddata
\end{deluxetable*}

\begin{figure*} 
\centering
\includegraphics[width=\textwidth]{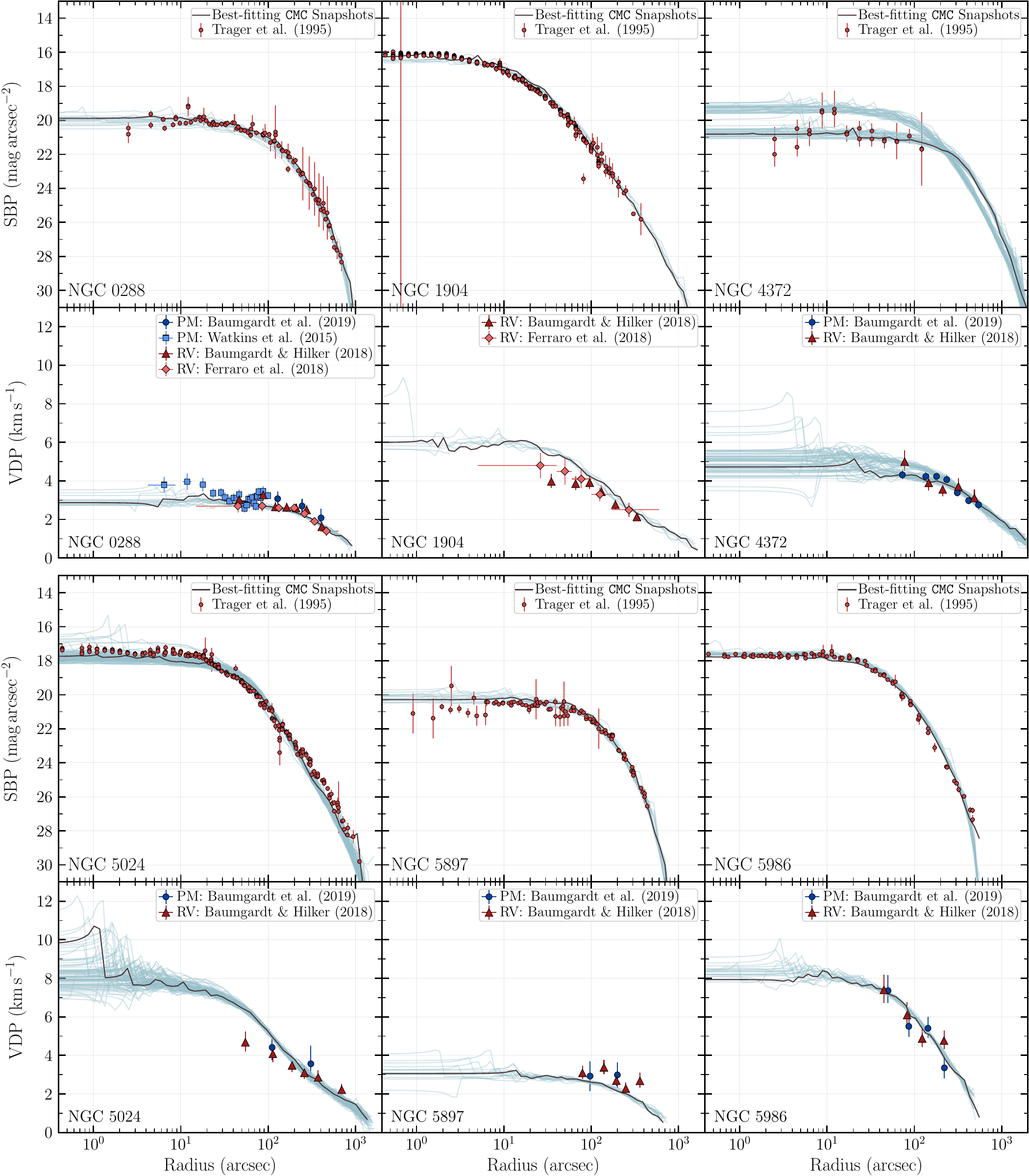}
\caption{The surface brightness and velocity dispersion profiles (SBPs and VDPs) of the MWGCs NGC~0288, NGC~1904, NGC~4372, NGC~5024, NGC~5897, and NGC~5986. The data points are the observed profiles, with SBPs from \cite{Trager1995} and VDPs compiled from both stellar proper motions \citep{Watkins2015,Baumgardt2019} and radial velocities \citep{Kamann2018,Ferraro2018,Baumgardt2018}. The gray curves correspond to model snapshots in the \texttt{CMC} Cluster Catalog \citep{kremer2020modeling} that fit well the observed profiles. We use the snapshots selected by \cite{Rui2021} that satisfy $s<10$ where $s=\max(\tilde{\chi}^2_{\rm SBP},\tilde{\chi}^2_{\rm VDP})$ is the maximum of the reduced $\chi^2$ statistics of the SBP and VDP fits (see their Table~5, Column~9). The solid black curves correspond to the single best-fitting snapshot, minimizing $s$.}
\label{fig:sbp_vdp_set_1}
\end{figure*}

\begin{figure*} 
\centering
\includegraphics[width=\textwidth]{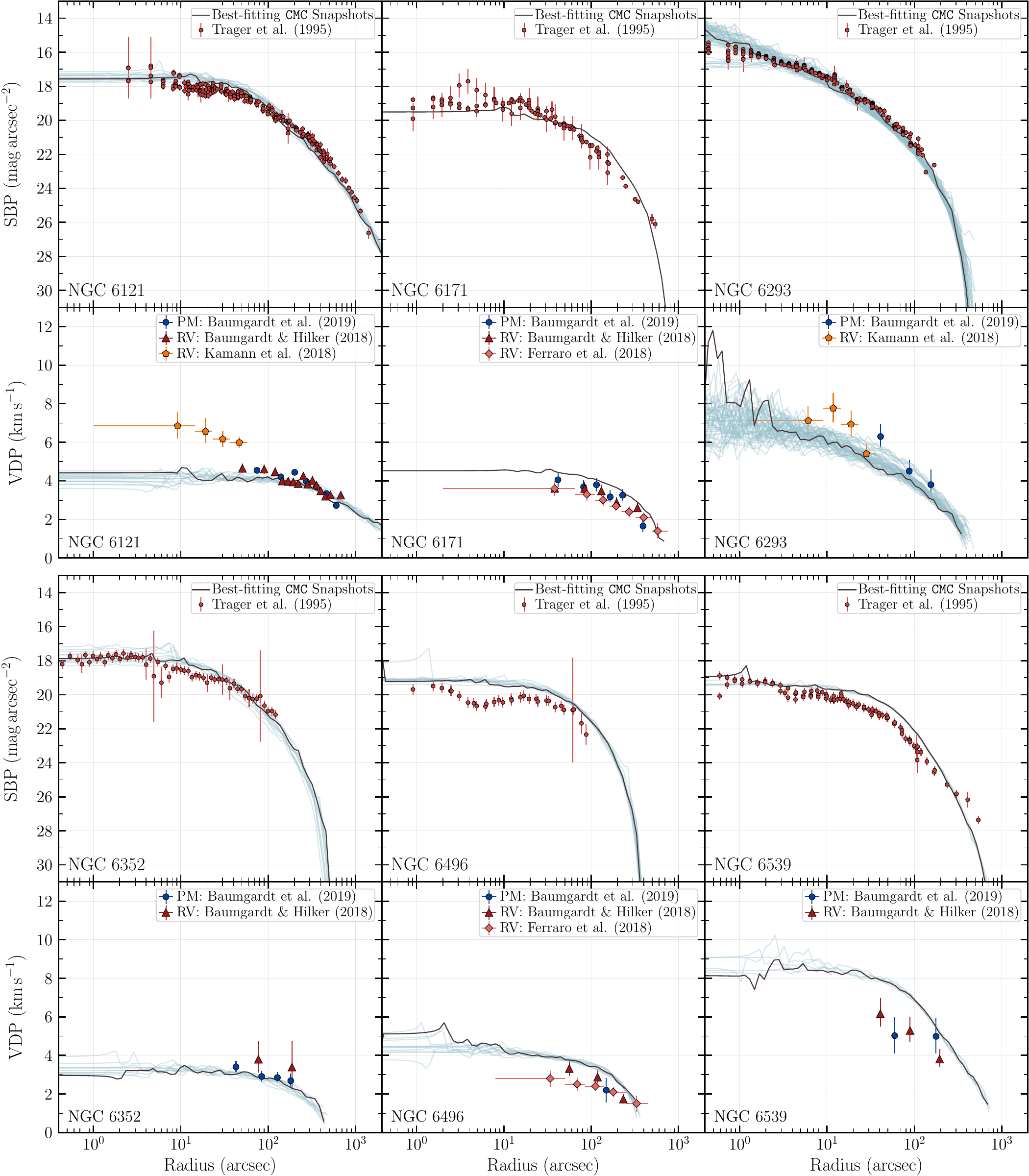}
\caption{As Figure~\ref{fig:sbp_vdp_set_1}, but for the MWGCs NGC~6121, NGC~6171, NGC~6293, NGC~6352, NGC~6496, and NGC~6539.}
\label{fig:sbp_vdp_set_2}
\end{figure*}

\begin{figure*} 
\centering
\includegraphics[width=\textwidth]{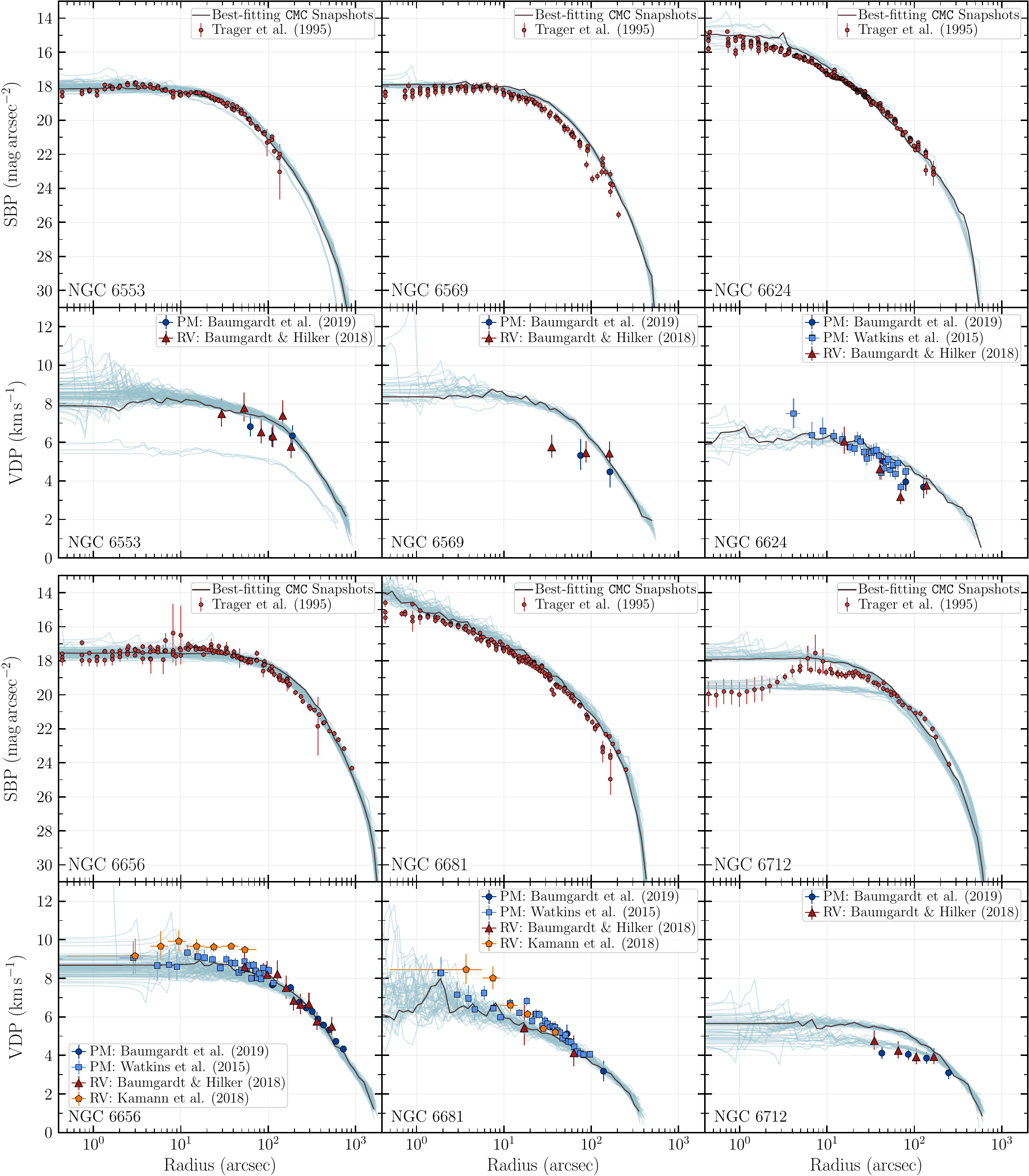}
\caption{As Figure~\ref{fig:sbp_vdp_set_1}, but for the MWGCs NGC~6553, NGC~6569, NGC~6624, NGC~6656, NGC~6681, and NGC~6712. For NGC~6624, the well-fitting \texttt{CMC} models come from \cite{Rui2021} rather than the raw \texttt{CMC} Cluster Catalog.}
\label{fig:sbp_vdp_set_3}
\end{figure*}

\begin{figure*} 
\centering
\includegraphics[width=\textwidth]{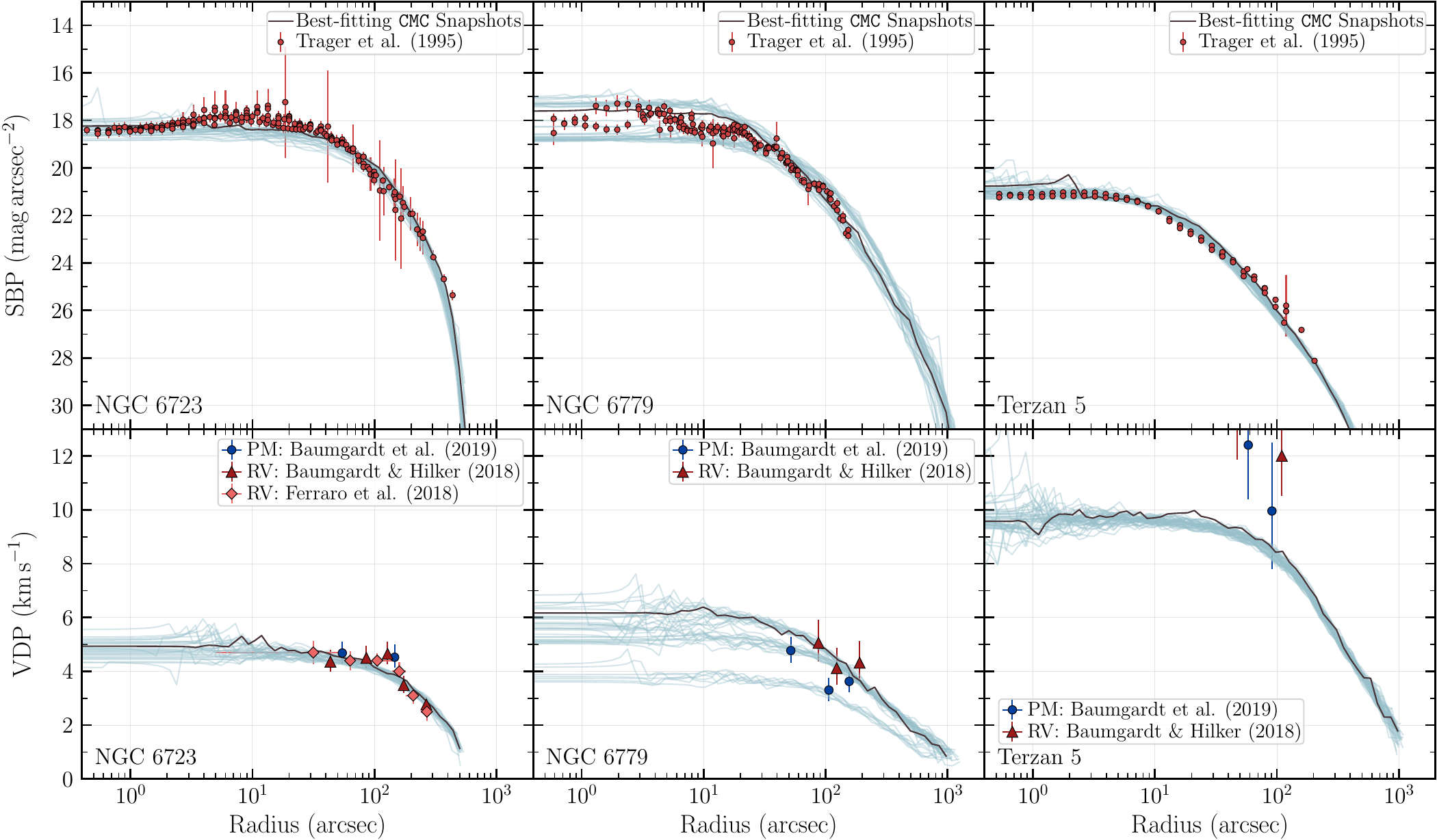}
\caption{As Figure~\ref{fig:sbp_vdp_set_1}, but for the MWGCs NGC~6723, NGC~6779, and Terzan~5.}
\label{fig:sbp_vdp_set_4}
\end{figure*}

\end{document}